\documentclass[prl,aps,superscriptaddress,floatfix,tightenlines,twocolumn]{revtex4}

\usepackage{amsmath}
\usepackage{amssymb}
\usepackage{amsfonts}
\usepackage{epsfig}
\usepackage{subfigure}
\usepackage[dvips]{color}
\usepackage{bm}
\usepackage{multirow}
\newcommand{\ket}[1]{|#1\rangle}       

\begin{document}

\title{Experimental testing of entropic uncertainty relations with multiple measurements in pure diamond}

\author{Jian Xing }
\affiliation{Beijing National Laboratory for Condensed Matter Physics,
Institute of Physics, Chinese Academy of Sciences, Beijing 100190, China}
\author{Yu-Ran Zhang}
\affiliation{Beijing National Laboratory for Condensed Matter Physics,
Institute of Physics, Chinese Academy of Sciences, Beijing 100190, China}
\author{Shang Liu}
\affiliation{School of Physics, Peking University, Beijing 100871, China}
\author{Yan-Chun Chang}
\affiliation{Beijing National Laboratory for Condensed Matter Physics,
Institute of Physics, Chinese Academy of Sciences, Beijing 100190, China}
\author{Jie-Dong Yue}
\affiliation{Beijing National Laboratory for Condensed Matter Physics,
Institute of Physics, Chinese Academy of Sciences, Beijing 100190, China}

\author{Heng Fan}
\email{hfan@iphy.ac.cn}
\affiliation{Beijing National Laboratory for Condensed Matter Physics,
Institute of Physics, Chinese Academy of Sciences, Beijing 100190, China}
\affiliation{Collaborative Innovation Center of Quantum Matter, Beijing 100190, China}
\author{Xin-Yu Pan}
\email{xypan@iphy.ac.cn}
\affiliation{Beijing National Laboratory for Condensed Matter Physics,
Institute of Physics, Chinese Academy of Sciences, Beijing 100190, China}
\affiliation{Collaborative Innovation Center of Quantum Matter, Beijing 100190, China}

\date{\today}
\begin{abstract}
One unique feature of quantum mechanics is the Heisenberg uncertainty principle, which
states that the outcomes of two incompatible measurements cannot simultaneously achieve
arbitrary precision.
In an information-theoretic context of quantum information, the uncertainty principle can be formulated
as entropic uncertainty relations with two measurements for a quantum bit (qubit) in
two-dimensional system. New entropic uncertainty relations are studied for
a higher-dimensional quantum state with multiple measurements, the uncertainty bounds
can be tighter than that expected from two measurements settings and cannot result
from qubits system with or without a quantum memory.
Here we report the first room-temperature experimental testing of the entropic
uncertainty relations with three measurements in a natural three-dimensional solid-state system:
the nitrogen-vacancy center in pure diamond.
The experimental results confirm the entropic uncertainty relations for multiple measurements.
Our result represents a more precise demonstrating of the fundamental uncertainty principle
of quantum mechanics.
\end{abstract}
\maketitle

One significant feature of quantum theory that differs from our everyday life experience is the
uncertainty principle. The uncertainty relation that bounds the uncertainties about the measurement outcomes
of two incompatible observables on one particle was first formulated by Heisenberg using the
standard deviation \cite{Heisenberg}. One widely accepted form of this relation
is expressed by the Heisenberg-Robertson relation \cite{HR}: $\Delta R\Delta S\geq|\langle[R,S]\rangle|/2$
where $\Delta R$ is the standard deviation of an observable $R$. Since this
form of relations is state-dependent on the left-hand-side, an improvement of uncertainty relation,
in an information-theoretic context, was subsequently proposed and expressed as
\cite{Kraus,Maassen} $H(R)+H(S)\geq\log_2[{1}/{c(R,S)}]$ where $H(R)$ denote and the Shannon
entropy of the probability distribution of the outcomes when $R$ is measured and
$c(R,S)\equiv\max_{j,k}|\langle r_j|s_k\rangle|^2$ given $|r_j\rangle$ and $|s_k\rangle$ the
eigenvectors of $R$ and $S$, respectively. In the presence of a quantum memory, the uncertainty relation can
be generalized as \cite{Berta} $H(R|\textrm{B})+H(S|\textrm{B})\geq\log_2[{1}/{c(R,S)}]+H(\textrm{A}|\textrm{B})$
where $H(R|\textrm{B})$ denotes the conditional von Neumann entropy.
It provides a bound on the uncertainties of the measurement outcomes depending on the
entanglement between measured particle $\textrm{A}$ and the quantum memory $\textrm{B}$
and is validated by recent experiments \cite{ex1,ex2}. These results as well as related investigations \cite{prl0,prl,prl2}
have been discovered to have many significant applications, such as the security proofs for quantum cryptography
\cite{QC,QC2}, nonlocality \cite{science} and the separability problem \cite{prl3}. Besides, in some recent researches, the
fundamental reason of uncertainty relations have been
investigated extending to more general theories such as thermodynamics \cite{H,Ren} and relativity \cite{Fengjun}.
It is indicated that the violation of the uncertainty relations would lead to a violation of the
second law of thermodynamics.

There are also efforts made to generalize the uncertainty relations to more than two observables \cite{review} and
the entropic uncertainty relations for multiple measurements with general condition are studied
theoretically by some of us \cite{Liu} and another group \cite{RPZ}.
The bounds \cite{Liu} for multiple measurements in higher-dimension are tighter than that obtained from two measurements results,
so those uncertainty relations provide a more precise description of the uncertainty principle
which may highlight the boundary between quantum and classical physics.
Besides, in principle,
the uncertainty relations are due to the superposition in one quantum system,
which differ from the well-studied nonlocality or entanglement in composite quantum systems.
Thus, the essence of those uncertainty relations can be well demonstrated in a
three-dimension quantum system like a spin-1 state for the reasons
that they cannot be obtained from the ordinary two measurements setting,
and it is an indivisible quantum system cannot resulting in non-locality or entanglement.

In this work, we report the first room-temperature proof-of-principle implementation of the entropic
uncertainty relations for multiple measurements via the generalized mutually unbiased inequality
\cite{Liu} in a solid-state system: the nitrogen-vacancy (NV) center in pure diamond single crystal.
An individual N-V center can be viewed as a basic unit of a quantum computer and is one
of the most promising candidates for quantum information processing (QIP), since many
coherent control and manipulation processes have been performed with this system
\cite{k1,k2,k3,k4,k5,k6,k7,k8,k9,k10,k11,k12,k13,k14,k15,k16}. Here, we demonstrate the
entropic uncertainty relations for multiple measurements via the triplet ground states of the
spin-1 electron spin of a NV centre. Since the entropic uncertainty relations is state dependent,
we further investigate different initial states of spin-1 electron spin of a NV center.
We also change the complementarity of three measured observables and
verify different types of entropic uncertainty relations for multiple measurements.
Moreover, our system is a truly three-level system and has
overcome the defects of post-selection in the most common optical systems,
which differs from earlier relative works.

\begin{figure}[t]
 \centering
\includegraphics[width=0.45\textwidth]{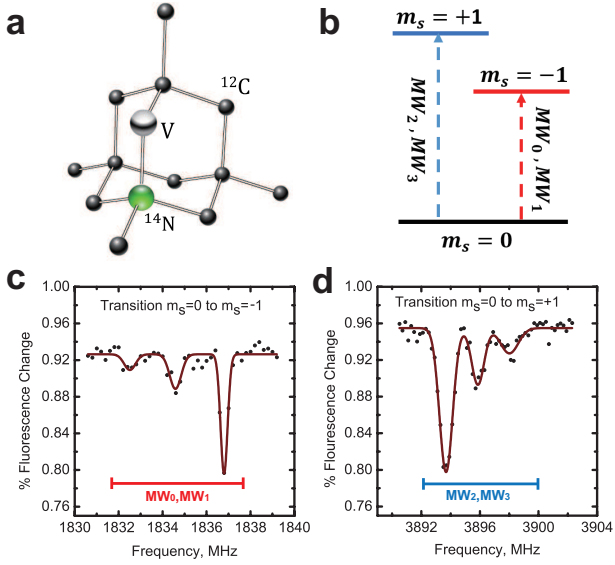}\\
\caption{\textbf{Typical structure of NV center in pure diamond single crystal.} (\textbf{a})
The NV center consists of a nearest-neighbor pair of a ${}^{14}$N atom, which substitutes for
a ${}^{12}$C atom, and a lattice vacancy (V). (\textbf{b}) Three energy levels of the ground state
of NV center. The electron spin state is controlled by MW pulses. MW$_0$
and MW$_2$ indicate MW pulses with a phase of $0$, while MW$_1$ and MW$_3$ indicate MW
pulses with a phase of $\pi/2$. (\textbf{c}) ODMR spectra of transition $m_s=0$ to $m_s=-1$.
(\textbf{d}) ODMR spectra of transition $m_{s}=0$ to $m_s=+1$.
}\label{fig:1}
\end{figure}
\section*{Results}

\textbf{System description and experimental setup.}
The electron spin of NV center interacts with the external magnetic field, causing a splitting of the three energy spin state.
The Hamiltonian of a negative charged NV center (NV$^-$) in pure
diamond under an external magnetic field $\bm{B}$ is written as
\begin{eqnarray}\label{eq:H}
H&=&\Delta S_{z}^2-\gamma_e \bm{B \cdot S}-\gamma_N\bm{B}\cdot \bm{I}^{(N)}-\gamma_c\bm{B}\cdot\sum_{i} \bm{I}_{i}^{(C)}\nonumber\\
&+&A_{\parallel}^{(N)} S_{z} I_{z}^{(N)}+A_{\perp}^{(N)} S_{x} I_{x}^{(N)}+A_{\perp}^{(N)} S_{y} I_{y}^{(N)}\nonumber\\
&+&S_z\sum_{i}\bm{A}_{i}\cdot\bm{I}_{i}^{(C)}
\end{eqnarray}
where $\Delta=2.87$~GHz is the zero-field splitting of the spin-1
ground states. $\gamma_e=1.76\times10^{11}$~T$^{-1}$s$^{-1}$  and
$\gamma_c=6.73\times10^7$~T$^{-1}$s$^{-1}$ are the gyromagnetic ratio
of electron spins and $^{13}$C nuclear spins.
$\bm{A}_{i}$ is the hyperfine tensor for $\bm{I}_{i}^{(C)}$.
A$_{\parallel}^{(N)}$ and A$_{\perp}^{(N)}$ are hyperfine constants for $\bm{I}^{(N)}$.
%
%
In this condition, the electron spin
couples with the I$_{N}=1$ (m$_{N_{s}}=\pm1,0$) nuclear spin, thus, m$_{s}=-1$ level splits into 3 energy
levels with states denoted by Dirac notation $|m_{N_{s}}, m_{s}\rangle$:
$|1,-1\rangle$, $|0,-1\rangle$ and $|-1,-1\rangle$.
Each one of the 3 transitions from a energy level that m$_{s}=0$ to another level with m$_{s}=-1$
indicates a dip in the spectra.

\begin{figure}[t]
 \centering
\includegraphics[width=0.47\textwidth]{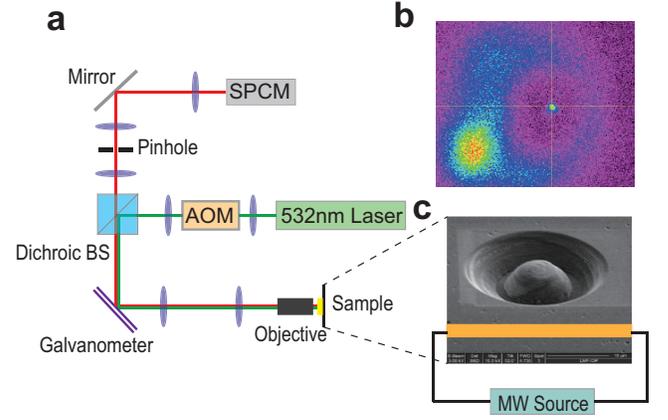}\\
\caption{\textbf{Experimental setup.} (\textbf{a}) Sketch map of the home-built scanning confocal microscope.
A 532 nm Laser beam from laser device is switched by an acoustic optic modulator (AOM) and focused on the
sample through a microscope objective. The fluorescence of NV center is collected by the same objective and
detected by the single photon count meter (SPCM). The galvanometer is used to perform an X-Y scan of the
sample while the dichroic beam-splitter (BS) is used to split the fluorescence of NV center and Laser. 
(\textbf{b}) Typical fluorescence scanning chart of the SIL and the NV center in it. (\textbf{c}) Typical photo 
of the SIL taken by electron microscope and sketch map of microwave system.
}\label{fig:2}
\end{figure}

The experiment is implemented with one single NV center in pure diamond (Sumitomo, Nitrogen Concentration $\ll$ 5~ppb).
The decoherence of NV centers in this sample is dominated by the nuclear spins of ${}^{13}$C atoms.  NV centers in diamond are surrounded by randomly distributed ${}^{13}$C atoms as the natural abundance of ${}^{13}$C is 1.1\%.
The
nuclear spin of ${}^{13}$C atom would interact with NV center electron spin leading to extra splitting and decoherence.
Typical dephasing time (${T}^{*}_{2}$) of NV centers in this sample is over 600 ns.
For a better manipulation fidelity, we choose an NV center without nearby ${}^{13}$C atom.
A permanent magnet is used to apply an external magnetic field on the system and is tuneable in both strength
and orientation.
Under several circumstance, excited-state level anti-crossing (ESLAC) of center electron spin is used for nuclear spin
polarization \cite{ESLAC_PRB}. When the magnetic field is around ESLAC point (about 507~Gauss), Laser driven electron spin
polarization would transfer nearby nuclear spins.
In this experiment, the magnet is adjusted to about 370~Gauss as the $^{14}$N nuclear spin is
partially polarized to improve the operation fidelity.

\begin{figure}[t]
 \centering
\includegraphics[width=0.45\textwidth]{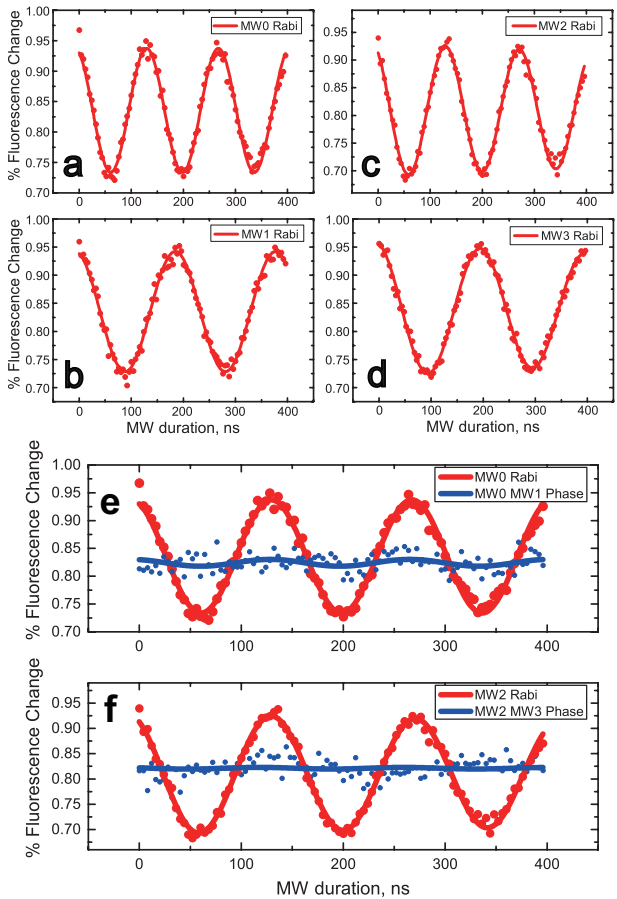}\\
\caption{\textbf{ Rabi oscillations carried out by the four MW channels.}
(\textbf{a}) MW$_0$. (\textbf{b}) MW$_1$. (\textbf{c}) MW$_2$. (\textbf{d}) MW$_3$.
(\textbf{e}) Red line shows the Rabi oscillation carried out by MW$_0$. Blue line shows the Rabi
oscillation carried out by MW$_1$ after a MW$_0$ $\frac{\pi}{2}$ pulse. (\textbf{f}) Red line shows the Rabi
oscillation carried out by MW$_2$. Blue line shows the Rabi oscillation carried out by MW$_3$
after a MW$_2$ $\frac{\pi}{2}$ pulse.
}\label{fig:3}
\end{figure}

Hyperfine spectra of the NV center
is obtained by optically detected magnetic resonance (ODMR) \cite{ODMR} scanning as shown is Fig.~\ref{fig:1}.
A home-built scanning confocal microscope combined with integrated microwave (MW) devices as shown in
Fig.~\ref{fig:2} is employed to initialize, manipulate and read out the electron spin state.
A 532 nm Laser beam from laser device is switched by an acoustic optic modulator (AOM) and focused on the
sample through a microscope objective. The fluorescence of NV center is collected by the same objective and
detected by the single photon count meter (SPCM). The galvanometer is used to perform an X-Y scan of the
sample while the dichroic beam splitter (BS) is used to split the
fluorescence of NV center and Laser.
Resonance microwave is used to control the electron spin state.
To enhance the photon collection efficiency, a solid immersion
lens (SIL) \cite{SIL_APL} is etched above an NV center. A coplanar waveguide (CPW) antenna is deposited
close to the SIL to deliver microwave pulses to the NV center.
Typical fluorescence scanning chart of the SIL and the NV center in it is shown in Fig.~\ref{fig:2}(b).
The photo of SIL taken by electron microscope and sketch map of microwave system is also indicated in Fig.~\ref{fig:2}(c).
Four MW channels (MW$_{0}$, MW$_{1}$,
MW$_{2}$, MW$_{3}$) are used which are controlled by individual RF switches for state and phase controls of
the electron spin (Fig.~\ref{fig:1}) in this experiment. MW$_{1}$ and MW$_{3}$ are respectively set to have a $\pi$/2 phase shift relative
to MW$_{0}$ and MW$_{2}$.  In Figs.~\ref{fig:3}(a-d), the Rabi oscillations carried out by the 
four MW channels are implemented.  Figs, 3(e) and 3(f) show that the relative phase between 
MW$_1$ and MW$_0$ and the relative phase between MW$_3$ and MW$_2$ are both $\frac{\pi}{2}$.


\begin{figure}[t]
 \centering
\includegraphics[width=0.45\textwidth]{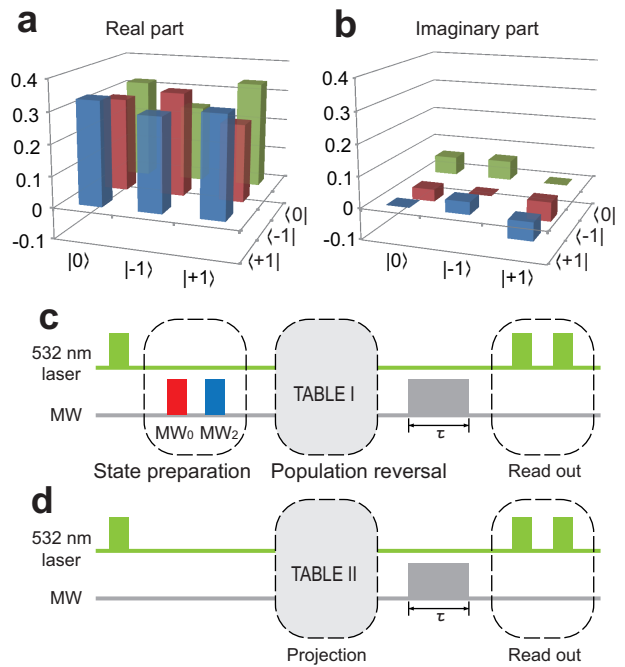}\\
\caption{\textbf{State tomography 
and pulse sequences for entropy measurement and state tomography.}
(\textbf{a}) Real part of state tomography result of an electron spin superposition state
$\frac{1}{\sqrt{3}}(|0\rangle+|-1\rangle+|+1\rangle)$. (\textbf{b}) Imaginary part of state
tomography result.
(\textbf{c}) Pulse sequence for state tomography. State preparation
is executed by adopting MW$_0$ with 26~ns and MW$_2$ with 26~ns. Population reversal is implemented
by MW pulses shown in TABLE~\ref{tab:2}.
(\textbf{d}) Pulse sequence for generalized entropic uncertainty relations for multiple measurements.
The projection scheme is carried out by MW pulses shown in TABLE~\ref{tab:1}. The MW pulse whose length is $\tau$
indicates the Rabi oscillation scheme.
}\label{fig:4}
\end{figure}

\begin{figure*}[t]
 \centering
\includegraphics[width=0.8\textwidth]{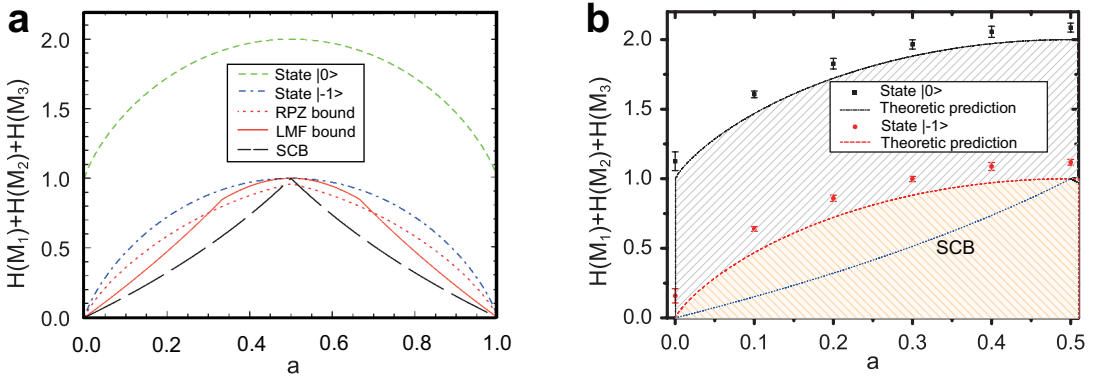}\\
\caption{\textbf{Entropic uncertainty relations for three measurements in the three-dimensional system.}
(\textbf{a}) Comparison between several bounds and entropic uncertainty with respect to $a$, including
the maximal SCB (long-dashed black line), RPZ bound (dotted red line) and LMF bound (solid orange line).
Dashed green line is for the theoretic result of state $|0\rangle$ and dashed-dotted blue line is for that
of state $|-1\rangle$. (\textbf{b}) Comparison between the predicted measurement entropy, experiment
results and  SCB with respect to parameter $a$. The error bars use the standard error (SE).}\label{fig:5}
\end{figure*}

To demonstrate that our system is a truly three-level system which can overcome the defects of
post-selection in common optical systems, we plots the state tomography result of an electron
spin superposition state $|\psi\rangle=\frac{1}{\sqrt{3}}(|0\rangle+|-1\rangle+|+1\rangle)$ in Fig.~\ref{fig:4}(a) and \ref{fig:4}(b).
See Methods for details.  Pulse sequence for state tomography is shown in Fig.~\ref{fig:4}(c) with MW
pulses for different measurement bases shown in TABLE~\ref{tab:2}. The fidelity
of the experimental result is about $95.35\%$, which is calculated from
$F(\rho)=\textrm{Tr}\sqrt{\sqrt{\sigma}{\rho}\sqrt{\sigma}}$ with $\sigma=|\psi\rangle\langle\psi|$.
As a result, our truly three-level system is well suitable for the investigation of  generalized entropic
uncertainty relations for multiple measurements.

\textbf{Entropic uncertainty for multiple relations and multiple measurements.}
Here we summarize the details of several multiple-measurement entropic uncertainty relations being used in the main article.
Generally, a multiple-measurement entropic uncertainty relation is of the following form.
\begin{equation}
\sum_{m=1}^NH(M_m)\geq B(M_1,M_2,...,M_N,\rho),
\end{equation}
where $\{M_m\}$ is a set of quantum measurements of cardinality $N$ and $B(\cdot)$ is some non-negative bound
which is generally a function of the measurements as well as the density operator $\rho$ of the measured system.

For experimental demonstration for entropic uncertainty relations for multiple measurements,
we choose to measure three measurement operators in three-dimensional space. Our system, a
truly three-level system, of which the quantum states corresponding to $m_s=0$, $m_s=-1$ and $m_s=+1$
are denoted by $|0\rangle$, $|-1\rangle$ and $|+1\rangle$, respectively.
Generally, the entropic uncertainty for the three-measurement
case is lower bounded by $B(M_1,M_2,M_3,\rho)$ which depends on the measurements $M_1$, $M_2$ and $M_{3}$
and chosen initial states $\rho$.
The measurements are chosen with eigenvectors as
\begin{eqnarray}
M_1&=&\{|0\rangle,|-1\rangle,|+1\rangle\},\label{e2}\\
M_2&=&\{\sqrt{0.5}(|0\rangle-|+1\rangle),|-1\rangle,{\sqrt{0.5}(|0\rangle+|+1\rangle)}\},\nonumber\\
M_3&=&\{\sqrt{a}|0\rangle+\sqrt{b}|-1\rangle,\sqrt{b}|0\rangle-\sqrt{a}|-1\rangle,|+1\rangle\},\nonumber
\end{eqnarray}
where $b=1-a$ and $a\in[0,1]$ is required.  For a detailed comparison, we take three different
lower bounds into consideration, which include Rudnicki-Puchala-Zyczkowski (RPZ) direct sum
majorization  bound \cite{RPZ}, simply constructed bound (SCB) and the recent generalized
Maassen-Uffink (MU) bound figured out by Liu, Mu and Fan (LMF) \cite{Liu}. See Methods for details.

The electron spin of NV center is initialized with a 532~nm laser pulse.
Projection measurements with three sets of eigenvectors are used to ensure the initial state of NV spin,
then the measurement entropy of each set of eigenvectors is determined. MW pulses of various length,
frequencies and phases as shown in Table I, are employed to carry out the projection. A Rabi oscillation
signal is used to read out the result after a projection. The pulse sequence is shown in Fig.~\ref{fig:4}(d).

Specifically, since entropic uncertainty relations are state dependent, we choose two initial
states $|0\rangle$ and $|-1\rangle$ in our experiment and the theoretic predictions compared
with the three kinds of lower bounds are shown in Fig.~\ref{fig:5}(a). It should be noticed that initial state $|-1\rangle$
is proven to have the minimum sum of entropies for the measurements expressed in Equations~(\ref{e2}).
The experimental results of the sum of entropies of two initial states with respect to different
values of $a$ are compared with the theoretic predictions in Fig.~\ref{fig:5}(b). These results have clearly verified
the entropic uncertainty relations predicted by the theory and the lower bounds. The difference between
the theoretic prediction and experiment result may be attributed to decoherence of electron spin
during the controls and measurements. Since the measured state is initially prepared as a pure state,
decoherence will increase the von Neumann of the measured state and enhance the sum of entropic
uncertainties. These analyzes can be also manifested by the lower bounds of entropic uncertainty
relations discussed in the Methods.

\section*{Disscusion}
In conclusion,  we report the first room-temperature implementation
of entropic uncertainty relations for three measurements in a three-dimensional solid-state system:
the nitrogen-vacancy center in pure diamond.  As summarized in Fig.~4b, we have experimentally investigated
entropic uncertainty relations for multiple measurements with different measured states
of spin-1 electron spin of a NV center and different kinds of three observables.
Differing from ordinary used optical systems, our system is a truly
three-level system and has overcome the defects of post-selection.
The significance of physics for multiple measurements is that
the uncertainty principle can be more precisely formulated
and demonstrated for a high-dimension quantum state.
Differring from the well-studied nonlocality, entanglement or other quantumness of correlations,
the uncertainty relations are due to the superposition principle in quantum mechanics.
Thus the demanding for physical implementation is that it should be an indivisible quantum system.
Our experiment system is naturally three-dimension, and our experimental results
confirm the theoretical expectation from the uncertainty relations.
Our result may shed new light on the differences between quantum and classical physics
in higher-dimension.

\section*{Methods}

\textbf{State tomography.}
State tomography is performed by projecting the initial state, denoted by $\rho$,
to three sets of eigenvectors indicated in Table~\ref{tab:2}. Fig.2(b) indicates the pulse sequence
of a state tomography measurement. The initial state is prepared by adopting MW$_{0}$
26~ns and MW$_{2}$ 26~ns to the electron spin of NV center, then Rabi oscillations carried
out by various MW channels are used to read out the projection value on each eigenvector
(Table II). Diagonal elements $\rho_{0,0}=\langle0|\rho|0\rangle$, $\rho_{-1,-1}=\langle-1|\rho|-1\rangle$,
$\rho_{+1,+1}=\langle+1|\rho|+1\rangle$ are obtained by projection values directly.
Non-diagonal elements, for example, $\rho_{-1,0}=\langle-1|\rho|0\rangle$ and
$\rho_{0,-1}=\langle0|\rho|-1\rangle$ are solved from a set of equations
\begin{eqnarray}
\langle\rho\rangle_{(|0\rangle-|-1\rangle)}&=& \rho_{0,0}+\rho_{-1,-1}-\rho_{0,-1}-\rho_{-1,0},\\
\langle\rho\rangle_{(|0\rangle-i|-1\rangle)} &=& \rho_{0,0} +\rho_{-1,-1}-i\rho_{0,-1}+i\rho_{-1,0}.
\end{eqnarray}
A $\pi$-pulse  of MW$_{2}$ is used to change the population between $m_{s}=0$ and $m_{s}=+1$ in order
to get the diagonal and nonagonal elements between $m_{s}=-1$ and $m_{s}=+1$.
The state tomography result of the electron spin superposition state
$\frac{1}{\sqrt{3}}(|0\rangle+|-1\rangle+|+1\rangle)$ is
\begin{equation}
\left[\begin{array}{ccc}
0.3314&0.2977-0.0392i&0.3200+0.0583i\\
0.2977+0.0392i&0.3306&0.2460+0.0621i\\
0.3200-0.0583i&0.2460-0.0621i&0.3380
\end{array}\right]
\end{equation}
with which von-Neumann entropy $S(\rho)=-\textrm{Tr}(\rho\log_2\rho)$ can be
calculated to be $0.4022$ and the fidelity is calculated.

\textbf{SCB.}
From the two-measurement MU inequality as well as the simple relation $H(M_i)\geq S(\rho)$, we can
easily obtain a lower bound as
\begin{eqnarray}
B&=&(N-\frac{k}{2})S(\rho)\nonumber\\
&-&\frac{1}{2}\log[c(M_1,M_2)c(M_2,M_3)...c(M_k,M_1)]
\end{eqnarray}
where we have $2\leq k\leq N$ or $k=0$ with which we define the second term of the r.h.s. to be zero.
We call the maximal value among all bounds deduced in this manner the \emph{simply constructed bound} (SCB),
which is explicitly expressed as
\begin{eqnarray}
B_\textrm{SCB}=\max_{k,\sigma}\left\{ -\frac{1}{2}\log[C_{k,\sigma}]
+\left(N-\frac{k}{2}\right)S(\rho)\right\},
\end{eqnarray}
where $C_{k,\sigma}:=c(M_{\sigma(1)},M_{\sigma(2)})...c(M_{\sigma(k)},M_{\sigma(1)})$.
Note that we have considered all possible permutations $\sigma$ among the indices of the measurements.

\textbf{LMF's generalized MU bound.}
In a recent work \cite{Liu} the following lower bound of generalized entropic uncertainty
relations for multiple measurements has been proven
\begin{equation}
B_\textrm{LMF}=(N-1)S(\rho)-\log(b),
\label{bound_MU}
\end{equation}
where
\begin{equation}
b=\max_{i_N}\left\{\sum_{i_2\sim i_{N-1}}\max_{i_1}[c(u^1_{i_1},u^2_{i_2})]\Pi_{m=2}^{N-1}c(u^m_{i_m},u^{m+1}_{i_{m+1}})\right\}.
\end{equation}
We regard this LMF lower bound as a generalization of the MU bound because it
explicitly reduces to MU bound if we take $N=2$.
%
One advantage of this result is that the role of the intrinsic uncertainty of the
pre-measurement state has been explicitly demonstrated.

\textbf{RPZ direct sum majorization bound.}
RPZ have introduced an alternative approach to multiple-measurement entropic uncertainty
relations in Ref.~\cite{RPZ}. 
Denote by $\ket{u^{(j)}_i}$ the $i$th basis vector of the $j$th measurement. By choosing a certain orthonormal
basis in the $d$-dimensional state space, we can rewrite all those basis vectors $\{\ket{u^{(j)}_i}\}$ as column
vectors in $\mathbb{C}^d$. Then define coefficients $\mathcal{S}_k$ as follow.
\begin{equation}
\mathcal{S}_k=\max\{\sigma_1^2(\ket{u^{(j_1)}_{i_1}},\ket{u^{(j_2)}_{i_2}},\cdots,\ket{u^{(j_{k+1})}_{i_{k+1}}})\},
\end{equation}
where $\sigma_1^2(\cdot)$ denotes the square of the largest singular value of a matrix and the maximum ranges over all \emph{subsets} $\{(i_1,j_1),(i_2,j_2),\cdots,(i_{k+1},j_{k+1})\}$ of cardinality $k+1$ of the set $\{1,2,\cdots,d\}\times\{1,2,\cdots,N\}$.
With this definition, a majorization relation as follows can be proven
\begin{equation}
\{p^{(j)}_i\}_{i,j=1}^{d,N}\prec \{\mathcal{S}_0,\mathcal{S}_1-\mathcal{S}_0,\mathcal{S}_2-\mathcal{S}_1,\cdots\},
\end{equation}
where $p^{(j)}_i$ is the probability of getting the $i$th outcome of the $j$th measurement.
This relation leads to the RPZ lower bound
\begin{equation}
B_\textrm{RPZ}=-\sum_{i=1}^{dN}(\mathcal{S}_i-\mathcal{S}_{i-1})\log(\mathcal{S}_i-\mathcal{S}_{i-1}).
\end{equation}



\begin{table}[b]
\begin{center}
\begin{ruledtabular}
\begin{tabular}{c| c c c}

 & Eigenvector & Population reversal & Rabi \\
\hline
\multirow{4}{*}{set 1}
& $|0\rangle$
& NA & MW$_{0}$\\
& $|-1\rangle$
& NA & MW$_{0}$\\
& $\sqrt{0.5}(|0\rangle-|-1\rangle)$
& NA & MW$_{0}$\\
& $\sqrt{0.5}(|0\rangle-i|-1\rangle)$
& NA & MW$_{1}$\\
\hline

\multirow{4}{*}{set 2}
& $|0\rangle$
& NA & MW$_{2}$\\
& $|+1\rangle$
& NA & MW$_{2}$\\
& $\sqrt{0.5}(|0\rangle-|+1\rangle)$
& NA & MW$_{2}$\\
& $\sqrt{0.5}(|0\rangle-i|+1\rangle)$
& NA & MW$_{3}$\\
\hline

\multirow{4}{*}{set 3} & $|0\rangle$
& MW$_{2}$ & MW$_{0}$\\
& $|-1\rangle$
& MW$_{2}$ & MW$_{0}$\\
& $\sqrt{0.5}(|+1\rangle-|-1\rangle)$
& MW$_{2}$ & MW$_{0}$\\
& $\sqrt{0.5}(|+1\rangle-i|-1\rangle)$
& MW$_{2}$ & MW$_{1}$\\

\end{tabular}
\end{ruledtabular}
\end{center}
\caption{\label{tab:2} \textbf{Eigenvectors for state tomography.} NA, not available.
A MW$_{2}$ $\pi$ pulse is used to carry out population reversal when eigenvector set3 is used. Rabi oscillation scheme is then executed by MW channel listed in collum "Rabi".}
\end{table}

\begin{table}[t]
\begin{center}
\begin{ruledtabular}
\begin{tabular}{c c c}
Eigenvector & MW channel & MW length \\
\hline
 (1 0 0)\footnote{The vector
$(\alpha,\beta,\gamma)$ stands for
 $\alpha|0\rangle+\beta|-1\rangle+\gamma|+1\rangle$.} & MW$_{0}$ & 0\\
 (0 1 0) & MW$_{0}$ & $\pi$\\
 (0 0 1) & MW$_{2}$ & $\pi$\\
(0 $\sqrt{0.5}$ $\sqrt{0.5}$) & MW$_{2}$, MW$_{0}$ & $\pi$, $1.5\pi$\\
(0 $\sqrt{0.5}$ $-\sqrt{0.5}$) & MW$_{2}$, MW$_{0}$ & $\pi$, $0.5\pi$\\
($\sqrt{0.1}$ $i\sqrt{0.9}$ 0) & MW$_{1}$ & $1.9\pi$\\
($\sqrt{0.9}$ $-i\sqrt{0.1}$ 0) & MW$_{1}$ & $0.1\pi$\\
($\sqrt{0.2}$ $i\sqrt{0.8}$ 0) & MW$_{1}$ & $1.8\pi$\\
($\sqrt{0.8}$ $-i\sqrt{0.2}$ 0) & MW$_{1}$ & $0.2\pi$\\
($\sqrt{0.3}$ $i\sqrt{0.7}$ 0) & MW$_{1}$ & $1.7\pi$\\
($\sqrt{0.7}$ $-i\sqrt{0.3}$ 0) & MW$_{1}$ & $0.3\pi$\\
($\sqrt{0.4}$ $i\sqrt{0.6}$ 0) & MW$_{1}$ & $1.6\pi$\\
($\sqrt{0.6}$ $-i\sqrt{0.4}$ 0) & MW$_{1}$ & $0.4\pi$\\
($\sqrt{0.5}$ $i\sqrt{0.5}$ 0) & MW$_{1}$ & $1.5\pi$\\
($\sqrt{0.5}$ $-i\sqrt{0.5}$ 0) & MW$_{1}$ & $0.5\pi$\\
($\sqrt{0.5}$ 0 $i\sqrt{0.5}$) & MW$_{2}$ & $1.5\pi$\\
($\sqrt{0.5}$ 0 $-i\sqrt{0.5}$) & MW$_{2}$ & $0.5\pi$\\
\end{tabular}
\end{ruledtabular}
\end{center}
\caption{\label{tab:1} \textbf{Projection of the Eigenvectors.} Each projection process is carried out by MW pulses from left to right with MW lengths listed behind.}
\end{table}

\end{document}